\documentclass[preprint,12pt,sort&compress]{elsarticle}
\usepackage{amsmath,amssymb,graphicx}
\usepackage{tabularx}
\newcolumntype{C}[1]{>{\centering\arraybackslash}p{#1}}
\usepackage{hyperref}
\usepackage[utf8]{inputenc}
\newcommand{\be}{\begin{equation}}
\newcommand{\ee}{\end{equation}}
\newcommand{\bea}{\begin{eqnarray}}
\newcommand{\eea}{\end{eqnarray}}
\newcommand{\non}{\nonumber}

\journal{Phys. Dark Univ.}

\begin{document}

\begin{frontmatter}

\title{Mimicking dark matter and dark energy in a mimetic model compatible with GW170817}

\author[unitn,tifpa]{Alessandro Casalino\corref{cor1}}
\ead{alessandro.casalino@unitn.it}

\author[unitn,tifpa]{Massimiliano Rinaldi\corref{cor1}}
\ead{massimiliano.rinaldi@unitn.it}

\author[unitn,tifpa]{Lorenzo Sebastiani\corref{cor1}}
\ead{lorenzo.sebastiani@unitn.it}

\author[okc,nordita]{Sunny Vagnozzi\corref{cor1}}
\ead{sunny.vagnozzi@fysik.su.se}\cortext[cor1]{Corresponding author}

\address[unitn]{Dipartimento di Fisica, Universit\`{a} di Trento,\\Via Sommarive 14, I-38123 Povo (TN), Italy}
\address[tifpa]{Trento Institute for Fundamental Physics and Applications (TIFPA)-INFN,\\Via Sommarive 14, I-38123 Povo (TN), Italy}
\address[okc]{The Oskar Klein Centre for Cosmoparticle Physics, Stockholm University, Roslagstullbacken 21A, SE-106 91 Stockholm, Sweden}
\address[nordita]{The Nordic Institute for Theoretical Physics (NORDITA), Roslagstullsbacken 23, SE-106 91 Stockholm, Sweden}

\begin{abstract}
\noindent The recent observation of the the gravitational wave event GW170817 and of its electromagnetic counterpart GRB170817A, from a binary neutron star merger, has established that the speed of gravitational waves deviates from the speed of light by less than one part in $10^{15}$. As a consequence, many extensions of General Relativity are inevitably ruled out. Among these we find the most relevant sectors of Horndeski gravity. In its original formulation, mimetic gravity is able to mimic cosmological dark matter, has tensorial perturbations that travel exactly at the speed of light but has vanishing scalar perturbations and this fact persists if we combine mimetic with Horndeski gravity. In this work, we show that implementing the mimetic gravity action with higher-order terms that break the Horndeski structure yields a cosmological model that satisfies the constraint on the speed of gravitational waves and mimics both dark energy and dark matter with a non-vanishing speed of sound. In this way, we are able to reproduce the $\Lambda$CDM cosmological model without introducing particle cold dark matter.
\end{abstract}


\end{frontmatter}

\section{Introduction}

The unknown fundamental nature of dark matter~\cite{Garrett:2010hd} and dark energy~\cite{Frieman:2008sn} opens the way to a theory of gravity that might differ from General Relativity (GR)~\cite{Nojiri:2006ri,Nojiri:2010wj,Clifton:2011jh,Capozziello:2011et,Nojiri:2017ncd}. In particular, failure to detect  different sorts of dark matter particles encourages the search for alternative scenarios. Similarly, the cosmological constant of the standard $\Lambda$CDM cosmological model suffers from a huge fine-tuning problem~\cite{Weinberg:1988cp} and several attempts are being made to overcome this issue by assuming a dynamical origin of the current accelerated expansion of the Universe~\cite{Copeland:2006wr,Bamba:2012cp}. In recent years, several models of modified gravity have been developed in order to reproduce the full expansionary history of our Universe without invoking the presence of forms of matter different other than standard baryonic matter and radiation~\cite{Nojiri:2006ri,Nojiri:2010wj,Clifton:2011jh,Capozziello:2011et,Nojiri:2017ncd}.

There are several different ways in which GR can be modified. For instance, one can add new combinations of the curvature invariants to the Hilbert-Einstein action (see for instance~\cite{Sotiriou:2008rp,DeFelice:2010aj,Nojiri:2010wj,Capozziello:2011et,Sebastiani:2015kfa} for the case of $F(R)$-gravity). Another approach is offered by scalar-tensor theories of gravity, wherein additional scalar fields are introduced and coupled to gravity. A particularly promising attempt in this direction is represented by Horndeski's theory of gravity~\cite{Horn}: this theory consists the most general class of scalar-tensor gravitational theories where the equations of motion are second-order differential equations (as in GR). Horndeski gravity provides a generic action avoiding Ostrogradski instabilities~\cite{Ostro1, Ostro2,Ostro3}. Aside from Horndeski gravity and $F(R)$ gravity, a plethora of other models have been explored. The possibility of taking some of these theories off the table will ultimately rely upon comparison with observations~\cite{Koyama:2015vza} (see e.g.~\cite{Alonso:2016suf} for the case of scalar-tensor gravity), and gravitational waves (GWs) represent an extremely promising arena in this direction.

Recently, the LIGO/Virgo collaboration observed the event GW170817, produced by merger of a binary neutron star system~\cite{TheLIGOScientific:2017qsa}. Thereafter, a number of counterparts across the electromagnetic (EM) spectrum were observed. In particular, the optical counterpart of the GW170817, the short gamma-ray burst event GRB170817A, was observed by the \textit{Fermi} Gamma-ray Burst Monitor and the Anti-Coincidence Shield on board the International Gamma-Ray Astrophysics Laboratory (\textit{INTEGRAL}) spectrometer~\cite{Monitor:2017mdv}. The association of GRB170817A to GW170817A, as well as the consistency of its arising from a binary neutron star merger, was confirmed in~\cite{Monitor:2017mdv,Goldstein:2017mmi,Murguia-Berthier:2017kkn}.

The optical counterpart GRB170817A was detected within a time-delay of $\delta t=(1.734 \pm 0.054)\,s$ from GW170817. Most of the time-delay is dominated by astrophysical contributions, associated to the collapse of the hypermassive neutron star formed during the merger. In~\cite{Monitor:2017mdv}, the size of these astrophysical contributions was conservatively estimated as $\approx 10\,s$. This implies that GWs travel at a speed $c_T$ which is extremely close to the speed of light: $c_T \approx 1$~\footnote{We use natural units wherein the speed of light is set to $c=1$.}. This astonishingly simple observation has already placed severe constraints on several theories of modified gravity: any modified gravity model predicting $c_T\neq 1$ must now be seriously reconsidered, and several previously viable theories of gravity are now excluded~\cite{Creminelli:2017sry,Sakstein:2017xjx,Ezquiaga:2017ekz,Baker:2017hug,Ezquiaga:2018btd} (see~\cite{Lombriser:2015sxa,Lombriser:2016yzn} for earlier important work, see also~\cite{Boran:2017rdn,Nojiri:2017hai,Arai:2017hxj,Amendola:2017orw,Visinelli:2017bny,Crisostomi:2017lbg,Langlois:2017dyl,
Gumrukcuoglu:2017ijh,Kreisch:2017uet,Bartolo:2017ibw,Dima:2017pwp,Cai:2018rzd,Pardo:2018ipy}). Still, one could  imagine models where the dispersion relation of GWs is modified such that $c_T=1$ today and for a range of wavelengths detectable by the LIGO/Virgo experiment. However, one might generically expect such a model to bring along severe fine-tuning problem. With these considerations in mind, in this paper we will focus on a model with $c_T^2=1$ throughout the entire evolutionary history and at all scales. The possibility of a tiny violation of the $c_T^{2} = 1$ constraint, within the uncertainty allowed by present data from the joint GW170817/GRB17081A detection, will be entertained in a companion paper~\cite{inprep}.

A particularly interesting theory of modified gravity is mimetic gravity, proposed in 2013 by Chamseddine and Mukhanov~\cite{m1} (see also~\cite{Lim:2010yk,Gao:2010gj,Capozziello:2010uv,Zumalacarregui:2013pma} for related and important earlier work). In the original work, the conformal degree of freedom of gravity was isolated in a covariant way, through a reparametrization of the physical metric $g_{\mu \nu}$ in terms of an auxiliary metric $\tilde{g}_{\mu \nu}$ and the mimetic scalar field $\phi$:
\begin{eqnarray}
g_{\mu \nu} = -\tilde{g}_{\mu \nu}\tilde{g}^{\alpha \beta}\partial_{\alpha}\phi\partial_{\beta}\phi \, .
\label{mimetic}
\end{eqnarray}
It is easy to show that, for consistency, the following condition has to be satisfied:
\begin{eqnarray}
g^{\mu \nu}\partial_{\mu}\phi\partial_{\nu}\phi = -1 \, .
\label{mimeticconstraint}
\end{eqnarray}
In~\cite{m1}, it was shown that the equations of motion resulting from the reparametrization of Eq.~(\ref{mimetic}) mimic a pressureless fluid on cosmological scales, which can identified with dark matter. Subsequently it was realized that the theory is related to GR via a non-invertible disformal transformation involving the mimetic field $\phi$, thus explaining why the dynamics of the theory are modified with respect to GR~\cite{Deruelle:2014zza,Domenech:2015tca,Achour:2016rkg}. A simple extension of the original model featuring a potential for the mimetic field, $V(\phi)$, has been shown to also be able to mimic dark energy and provide an early-time inflationary era, as well as some allowing for bouncing solutions~\cite{m2}. In~\cite{m2}, it was also argued that the mimetic constraint Eq.~(\ref{mimeticconstraint}) can be enforced at the level of the action through a Lagrange multiplier term. An incomplete list of works examining astrophysical and cosmological issues in mimetic gravity can be found in~\cite{Golovnev:2013jxa,Barvinsky:2013mea,Nojiri:2014zqa,Saadi:2014jfa,Capela:2014xta,Mirzagholi:2014ifa,Leon:2014yua,
Haghani:2015iva,Matsumoto:2015wja,Momeni:2015gka,Myrzakulov:2015sea,Astashenok:2015haa,Myrzakulov:2015qaa,
Myrzakulov:2015kda,Odintsov:2015wwp,Hammer:2015pcx,Ramazanov:2016xhp,Nojiri:2016ppu,Nojiri:2016vhu,Sadeghnezhad:2017hmr,
Baffou:2017pao,Vagnozzi:2017ilo,Bouhmadi-Lopez:2017lbx,Shen:2017rya,Nojiri:2017ygt,Dutta:2017fjw,Golovnev:2018icm,Langlois:2018jdg,
Brahma:2018dwx,deHaro:2018sqw,Zhong:2018tqn,Chamseddine:2018qym,Chamseddine:2018gqh,Zlosnik:2018qvg,Nashed:2018qag}, while a recent review can be found in~\cite{vase}.

A comment on local gravity tests of mimetic gravity is in order. We notice that, to date, there is no work thoroughly examining constraints on mimetic gravity from local gravity, i.e. from modifications to the Newtonian potential within the Solar System. Efforts towards this direction were nonetheless carried out in~\cite{Babichev:2016jzg}, which studied local gravity constraints within a mimetic model closely related to the one we will study in this work [i.e. setting $a=b=0$ in the action given by Eq.~(\ref{lagrangian})]. By studying the precession of Mercury's perihelion, in~\cite{Babichev:2016jzg} the constraint $c \lesssim 10^{-18}$, where $c$ is the Lagrangian parameter appearing in the model we will study [Eq.~(\ref{lagrangian})], was found. In general, however, we note that mimetic models suffer from caustic instabilities on small (galactic and subgalactic) scales, which imply that in order for the theory to be mathematically consistent it requires an appropriate ``completion" which removes these caustic instabilities. So far, the issue of how to remove the caustic instabilities has yet to receive a definite solution, although work in this direction has been pursued~\cite{Capela:2014xta,Babichev:2016jzg,Babichev:2017lrx}. Nonetheless, as noted in~\cite{Babichev:2016jzg}, the details of how the mimetic theory is ``completed" to avoid caustic instabilities on small scales will then inevitably affect conclusions concerning local gravity constraints. In the absence of a compelling caustic-free mimetic theory which would lead to equally compelling local gravity constraints, herein we conservatively choose to refrain from discussing local gravity constraints on mimetic gravity further. The issue of local constraints on mimetic gravity, nonetheless, is admittedly an important open problem to which it is definitely worth returning within a more thorough study, which however falls beyond the scope of our work.~\footnote{We further note that no work so far has examined the possibility of invoking screening mechanisms to evade local gravity constraints if necessary. Vainshtein-like screening mechanisms generally require non-linear kinetic terms which are likely to be strongly constrained following the joint GW170817/GRB170817A detection. It is moreover conceivable that it would be in any case hard to implement screening mechanisms in mimetic gravity given that, at least in the original scenario, the mimetic field is non-dynamical and constrained (see however also~\cite{Ganz:2018vzg} examining the gravitational slip in mimetic theories consistent with GW170817/GRB170817A). We choose therefore not to discuss this issue further in our paper.}

A particularly appealing variant of the original mimetic theory starts from a ``seed'' Horndeski action rather than the Einstein-Hilbert one: in other words, the mimetic constraint Eq.~(\ref{mimeticconstraint}) is enforced on the scalar degree of freedom of Horndeski gravity through a Lagrange multiplier term in the action. The resulting mimetic Horndeski theory has been proposed in~\cite{Arroja:2015wpa} and subsequently studied in e.g.~\cite{Rabochaya:2015haa,Arroja:2015yvd,Cognola:2016gjy,Arroja:2017msd}. On a cosmological background, the theory features a fluid mimicking dark matter. However, at the perturbative level, this theory features some problems. In fact, the mimetic constraint kills the wave-like parts of the Horndeski scalar degree of freedom and removes the scalar degree of freedom of the theory. This implies that the speed of scalar perturbations (the sound speed $c_s$) vanishes. It is worth clarifying that a vanishing sound speed is problematic only if one wishes to perform inflation with the mimetic field, because the resulting perturbations would fail in explaining structure formation, as explained in~\cite{m2}. In fact, if $c_s=0$, perturbations of the mimetic field do not propagate in space. Quantizing such a field, and consequently generating vacuum quantum fluctuations, is problematic for many reasons (for instance, it would be hard to satisfy the appropriate commutation relation with the conjugate momentum). This in turn hinders the generation of perturbations which will then grow under gravitational instability to form the large-scale structure, one of the most important outcomes of successful inflation. Perhaps more importantly, a vanishing sound speed of the inflaton is also problematic from the observational point of view. In fact, measurements of the CMB temperature and polarization anisotropies from the \textit{Planck} satellite (and in particular the absence of detection of primordial non-Gaussianity) favour a speed of sound for the inflaton $c_s=1$, with a 95\% confidence level lower bound of $c_s > 0.024$~\cite{Ade:2015lrj}. This result excludes $c_s=0$ at high significance~\cite{Ade:2015lrj}. We wish to stress nonetheless that a vanishing speed of sound is strictly speaking only a problem if one wishes to perform inflation with the mimetic field, and not if one is only aiming at describing dark matter (for which $c_s=0$ is instead quite natural, although a very tiny but non-zero sound speed could nonetheless be desirable in order to possibly avoid caustic instabilities~\cite{Capela:2014xta,Babichev:2016jzg,Babichev:2017lrx}).

At any rate, it is worth considering modifications to the original mimetic scenario which allow for a non-vanishing sound speed. An obvious way to address this issue is to break the Horndeski structure of the theory, and thereby removing the special tuning guaranteeing that the equations of motion are at most of second order. Nonetheless, the presence of the mimetic constraint prevents the appearance of higher-than-second-order derivatives in the equations of motion. In this work, we shall follow this procedure, and consider the mimetic model proposed in~\cite{Cognola:2016gjy}, obtained by breaking the Horndeski structure of a starting mimetic Horndeski model, thus allowing for a non-zero sound speed. The model is theoretically appealing as it appears in the low-energy limit of projectable Ho\v{r}ava-Lifshitz gravity, a well-motivated candidate theory of quantum gravity. We will show that within this model, after imposing constraints on the speed of GWs arising from the detection of GW170817/GRB170817A, it is possible to mimic the $\Lambda$CDM evolutionary history wherein the Universe is filled with dark matter in agreement with observations.

This paper is organized as follows. In Sec.~\ref{sec:background} we define the action of the mimetic model we consider, and derive its equations of motion on a flat FLRW background. In Sec.~\ref{sec:perturbations}, we then perturb the FLRW line-element, first considering scalar perturbations (Subsec.~\ref{subsec:scalar}) which allow us to derive the sound speed $c_s$, and subsequently tensor perturbations (Subsec.~\ref{subsec:tensor}) which allow us to derive the gravitational wave speed $c_T$ and hence consider constraints on the parameters of the model from the joint GW170817/ GRB170817A detection.   In Subsec.\ \ref{subsec:stability} we address the problem of gradient and ghost instabilities of the theory. In Sec.~\ref{sec:late_time} we then consider late-time solutions which mimic dark matter and dark energy in agreement with observations, first in a simplified vacuum case (Subsec.~\ref{subsec:vacuum}), and then in a realistic setting adding radiation and baryonic matter (Subsec.~\ref{subsec:adding}), and finally calculate the resulting age of the Universe. We summarize our main findings and provide concluding remarks in Sec.~\ref{sec:conclusions}.

\section{Background equations}
\label{sec:background}

\noindent We consider the mimetic theory defined by the following action~\footnote{See also~\cite{Rinaldi:2016oqp,Diez-Tejedor:2018fue,Casalino:2018mna}, where related Horndeski and beyond Horndeski models, as well as their ability to mimic dark matter on cosmological and galactic scales, were studied.}:
\bea\label{lagrangian}\non
S&=& \int d^4 x \,\sqrt{-g} \Big[ R(1+2aX) -{c\over 2}(\square\phi)^{2}+{b\over 2}(\nabla_{\mu}\nabla_{\nu}\phi)^{2}-{\lambda\over 2}(2X+1) \\
&-&V+ \mathcal{L}_m \Big]\,,
\eea
where we set $16\pi G_N=1$ ($G_N$ is Newton's constant), $g$ is the determinant of the metric tensor $g_{\mu\nu}(x^i)$, $\mathcal{L}_m$ is the action of standard matter and radiation, $\phi$ is the mimetic field, $V\equiv V(\phi)$ is a potential for the mimetic field, and $X\equiv(1/2) g_{\mu\nu}\nabla^{\mu}\phi\nabla^{\nu}\phi$ is the kinetic term of the field. The Lagrangian multiplier $\lambda$ is introduced to enforce the mimetic constraint Eq.~(\ref{mimeticconstraint}) on the mimetic field, while $a\,,b\,,c$ are constant parameters. Note that when $b=c=4a$ we recover a mimetic Horndeski model~\cite{Horn}. Breakage of the Horndeski structure of the action is necessary in order for scalar perturbations to propagate ~\cite{Arroja:2015yvd,Cognola:2016gjy}. However, we will see that, on a Friedmann-Lema\^itre-Robertson-Walker (FLRW) background, the model  preserves the solutions of the corresponding mimetic Horndeski Lagrangian up to a (constant) rescaling of the effective Plank mass of the theory.

The action in Eq.~(\ref{lagrangian}) was first considered in~\cite{Cognola:2016gjy}, by explicitly breaking the Horndeski structure of a starting mimetic Horndeski model. In~\cite{Cognola:2016gjy}, it was argued that the model is related to the low-energy limit of Ho\v{r}ava-Lifshitz gravity~\cite{Horava:2009uw}, a candidate theory of quantum gravity which achieves power-counting renormalizability by explicitly breaking diffeomorphism invariance. In fact, the motivation for studying the original mimetic Horndeski model in~\cite{Cognola:2016gjy} was related to attempts to achieve power-counting renormalizability \`{a} la Ho\v{r}ava, albeit via a dynamical rather than explicit breaking of diffeomorphism invariance, through a non-standard coupling of curvature to the energy-momentum tensor of an exotic fluid. In these ``covariant renormalizable gravity'' theories, issues of infrared strong-coupling in Ho\v{r}ava-Lifshitz gravity, due to the appearance of an unphysical extra mode related to the explicit breaking of diffeomorphism invariance~\cite{Blas:2009yd,Blas:2009qj,Blas:2009ck}, are circumvented.

Let us now consider the equations of motion of our mimetic model, which are obtained by varying the action with respect to the metric, the Lagrange multiplier, and the mimetic field. Varying the action with respect to the metric we get
\bea\label{einstein_equations}\non
&&(1+2aX)G_{\mu\nu}+\nabla_{\mu}\phi\nabla_{\nu}\phi\left(aR-{\lambda\over 2}\right)\\\non
&&-\frac12 g_{\mu\nu}\left[ {b\over 2}\phi_{\alpha\beta}\phi^{\alpha\beta} -{c\over 2}(\square \phi)^{2}-{\lambda\over 2}(2X+1)-V  \right]\\\non
&&+2a(g_{\mu\nu}\square X-\nabla_{\mu}\nabla_{\nu}X)\\\non
&&-{b\over 2}g^{\alpha\beta}\left[ \nabla_{\alpha}(\phi_{\mu\beta}\nabla_{\nu}\phi)+
\nabla_{\alpha}(\phi_{\nu\beta}\nabla_{\mu}\phi)-\nabla_{\alpha}(\phi_{\mu\nu}\nabla_{\beta}\phi) \right]+b\phi^{\alpha}_{\,\,\,\mu}\phi_{\alpha\nu}\\
&&+{c\over 2}\left[ \nabla_{\nu}\phi\nabla_{\mu}(\square \phi) +\nabla_{\mu}\phi\nabla_{\nu}(\square \phi)-g_{\mu\nu}g^{\alpha\beta}\nabla_{\alpha}(\square \phi\nabla_{\beta}\phi) \right]=\frac{1}{2} T_{\mu \nu}\,,
\eea
where $G_{\mu\nu}$ is Einstein's tensor and $T_{\mu\nu}$ represents the stress tensor of standard baryonic matter and radiation. 
Here we set $\phi_{\mu\nu}\equiv\nabla_{\nu}\nabla_{\mu}\phi$. Variation of the action with respect to the field $\phi$ yields:
\begin{align}
& (\lambda-2 a R) (\nabla_\mu \nabla^\mu\phi) + (\nabla_\mu \phi)(\nabla^\mu \lambda) - 2 a (\nabla_\mu R) (\nabla^\mu \phi)\non\\
&+ b (\nabla_\nu \nabla_\mu \nabla^\nu \nabla^\mu \phi) - c (\nabla_\mu \nabla^\mu \nabla_\nu \nabla^\nu \phi) - \frac{\partial V(\phi)}{\partial \phi}=0\,.
\label{phigen}
\end{align}
Finally, variation with respect to the Lagrange multiplier $\lambda$ enforces the mimetic constraint:
\bea
X=-\frac12\,.
\eea 

We choose to work within a flat FLRW space-time, whose line-element is given by:
\begin{equation}
ds^2=-dt^2+A(t)^2(dx^2+dy^2+dz^2)\,, 
\end{equation}
In the above, $A\equiv A(t)$ is the cosmological scale factor and is a function of the time only. We have chosen to denote the scale factor by $A$ instead of $a$ in order to avoid possible confusion with the Lagrangian parameter $a$ in Eq.~(\ref{lagrangian}). On this background, the mimetic constraint immediately allows for the identification 
of the field with the cosmological time (up to a constant), namely:
\bea\label{mim_phi}
\phi=t\,.
\eea
With this identification, from the $(0,0)$ and $(1,1)$ components of (\ref{einstein_equations}) we obtain the equations of motion (EOMs):
\begin{align}
&(-6c+24a)\dot H+(36a+9c+12-9b)H^2-2V-2\lambda -2\rho_{\rm m}=0\,,\label{ttcomp}\\
&3H^2+2\dot H={2V-2P_{\rm m}\over 4-4a-b+3c}\,,\label{final}
\end{align}
where $H\equiv H(t)=\dot A/A$ is the Hubble parameter. Here, we denote the time derivative with a dot, and $\rho_{\rm m}$ and $P_{\rm m}$ correspond to the combined energy density and pressure of baryonic matter and radiation. 

The Klein-Gordon (KG) equation of the field, Eq.~(\ref{phigen}), taking into account Eq.~(\ref{mim_phi}), reads:
\bea\label{eomlambda_equiv}
\frac{1}{A^3} \frac{d}{d t} \left[A^3 \left(\lambda + (3b - 24a) H^2 + (3c - 12a)\dot{H}\right)\right] = - \frac{dV}{dt}\,,
\eea
while the continuity equation $\nabla_{\mu}T^{\mu\nu}=0$ for baryonic matter and radiation assumes the standard form:%
\begin{equation}\label{continuity_matter}
\dot\rho_{\rm m}+3H(\rho_{\rm m}+P_{\rm m})=0\,. 
\end{equation}
Note that when the Horndeski structure for the $\phi$ sector of the action Eq.~(\ref{lagrangian}) is recovered, i.e. when $c=4a$, the $\dot{H}$ term in Eq.~(\ref{eomlambda_equiv}) disappears, leaving a second order differential equation. 

Given the equation of state of the matter fluid (\ref{continuity_matter}), once the form of the potential $V$ is chosen, the system of equations (\ref{ttcomp},\ref{final}) can be solved with respect to $A(t)$ and $\lambda$. Alternatively, one can use Eq.~(\ref{eomlambda_equiv}) with one of Eqs.~(\ref{ttcomp}, \ref{final}). In this case, from Eq.~(\ref{eomlambda_equiv}) we get:
\begin{equation}
\rho_\text{df} (t) = \frac{C}{A(t)^3}+\frac{3}{A(t)^3}\int^t V(t') A(t')^3 H(t') dt'\,,\label{rhodf}
\end{equation}
where $C>0$ is an integration constant which sets the amount of mimetic dark matter, as the corresponding contribution to the energy density decays as $a^{-3}$, as expected for a pressureless component. In the above expression, $\rho_\text{df}$ is defined by:
\bea
\rho_\text{df}:=\lambda+V+(3 b - 24 a) H^2 + (3 c - 12 a)\dot H\label{rhoP}\,,
\eea
and can be read as an effective energy density of an induced \textit{dark fluid} with effective pressure 
\bea 
 P_\text{df}:=-V\,.\label{PP}
\eea
It is then easy to verify that:
\bea\label{continuity_df}
\dot\rho_{\text{df}} + 3 H (\rho_\text{df} + P_\text{df}) = 0\,.
\eea
Rearranging Eqs.~(\ref{ttcomp})-(\ref{final}) we obtain:
\begin{eqnarray}\label{friedmann}
6H^2 &=&\frac{4}{4-b + 3c - 4a}\, \left(\rho_\text{df} + \rho_\text{m}\right) \,,\nonumber\\
-4\dot H - 6H^2&=&\frac{4}{4-b + 3c - 4a} \left(P_\text{df} + P_\text{m}\right)\,.
\end{eqnarray}
We recognize the above as being Friedmann-like equations, with the Planck mass rescaled by a factor $(4-b+3c-4a)/4$. The quantity by which the Planck mass is rescaled determines the effective Newton constant. Enforcing that the rescaling is positive implies:
\begin{equation}
4-b+3c-4a>0\,.\label{condizione}
\end{equation}
Notice that for $a=b=0$ we recover the results of~\cite{m2}, which extended the original mimetic action by a term proportional to $(\Box \phi)^2$.

We immediately see that a constant potential $V$ in Eqs.~(\ref{rhoP},\ref{PP}) can be exploited to model dark matter and dark energy through the corresponding fluid. For more complex potentials, given in the action as functions of $\phi$ (and therefore of $t$ thanks to the mimetic constraint) and not as functions of the scale factor $A$, the dark fluid will model various types of fluids while leaving the dark matter sector unchanged.

In the next sections we will explore the perturbations of this model and find physical constraints on the Lagrangian parameters. Then we will analyse the background solutions for different forms of the potential $V$.

\section{Perturbations on a FLRW background}
\label{sec:perturbations}

\noindent As mentioned above, one of the main problems in mimetic gravity is the vanishing sound speed, implying the non-propagation of scalar perturbations. The problem persists even in mimetic Horndeski gravity. We have seen that breaking the Horndeski form of the $\phi$ sector, we can find a non-vanishing sound speed~\cite{Arroja:2015yvd,Cognola:2016gjy}. However, this comes at the risk of modifying the speed of gravitational waves $c_T$, which in the original mimetic gravity model is identically equivalent to the speed of light, $c_T=1$. Enforcing that $c_T$ remains equal to the speed of light when considering the mimetic model of Eq.~(\ref{lagrangian}) will strongly constrain the Lagrangian parameters. We will now discuss these issues in detail, and begin by computing the sound speed $c_s$ and the gravitational wave speed $c_T$.

\subsection{Scalar perturbations}
\label{subsec:scalar}

We begin by considering scalar perturbations around a flat FLRW line-element. In Newtonian gauge, the perturbed metric reads:
\bea
ds^{2}=-\left[1+2\Phi(t,x,y,z)\right]dt^{2}+A(t)^{2}\left[1-2\Psi(t,x,y,z)\right]\delta_{ij}dx^{i}dx^{j}\,.
\eea
We perturb the mimetic field and the Lagrange multiplier field as follows:
\bea
\phi=t+\delta\phi(t,x,y,z)\,,\quad \lambda=\lambda_{0}(t)+\lambda_{1}(t,x,y,x)\,,
\eea
where $|\delta\phi/t|\,,|\lambda_1/\lambda|\ll 1$. The mimetic constraint yields $\Phi=\delta\dot\phi$ and the $i\neq j$ components of the perturbed field equations [Eq.~(\ref{einstein_equations})] give:
\bea
(1-a)\Psi-\frac{b}{2}H\delta\phi+\left( a-\frac{b}{2}-1\right)\delta\dot\phi=0\,.
\eea
By substituting this result into any of the $tj$ components of Eq.~(\ref{einstein_equations}) leads to:
\bea\label{spert}\non
\delta\ddot\phi+H\delta\dot\phi+\left[\dot H + \frac{c_s^2 (\rho_{\rm m} + P_\text{m})}{b-c}\right]
\delta\phi-{c_{s}^{2}\over A^{2}}\nabla^{2}\delta\phi= -\frac{A(t) c_s^2}{b-c} (\rho_{\rm m} + P_\text{m}) v_{\rm m}\,,\\
\eea
where 
\bea\label{sound}
c_{s}^{2}={ 2(b-c)(a-1)\over (2a-b-2)(4-4a-b+3c)}\,,
\eea
is the squared speed of sound. In the expression above, $v_{\rm m}$ is the matter velocity. 

Recall that we have defined $\rho_{\rm m}$ and $P_{\rm m}$ to include both the baryonic matter and radiation components, although in principle one could separated them in the above discussion: that is, the term $(p_\text{m} + \rho_\text{m}) v_\text{m}$ should really be considered as a sum over the baryonic matter and radiation contributions. Notice also that in the limit $a=0, b=0$ and $c=-2\gamma$ we find the results of~\cite{m2}. Notice finally that these results are independent of the choice of the field potential $V$.

\subsection{Tensor perturbations}
\label{subsec:tensor}

Let us now turn our attention to tensor perturbations. The line-element is perturbed as:
\bea\non
ds^2 &=& -dt^2 +A(t)^2(1+h_{+})dx^2+2A(t)^2h_{\times}dxdy+A(t)^2(1-h_{+})dy^2\\
&+&A(t)^2dz^2\,, 
\eea
where we have chosen the TT-gauge. We denote by $h_{\times,+}$  the two polarisation states of the linear tensor perturbations. By inserting this into the Einstein equations and by using the unperturbed equations, we find the perturbed equation at the first order, where $h=h_{\times}$ or $h=h_{+}$:
\bea\label{tpert}
\ddot h+3H\dot h-{2(1-a)\over 2-2a+b}{1\over A(t)^{2}}{\partial^{2}h\over \partial z^{2}}=0\,.
\eea
From the above, we read off the squared gravitational wave speed 
\bea\label{tensor}
c_{T}^{2}={2(1-a)\over 2-2a+b}\,.
\eea
Clearly, the gravitational wave speed is in general different from the speed of light. Notice furthermore that the tensor perturbations are not affected by the presence of standard matter.

We clearly see that in order to satisfy the recent constraint from GW170817 /GRB17081A which enforces $c_T^2 \simeq 1$, we have to consider $\vert b \vert \ll 1$. In fact, the requirement that $c_T \equiv 1$ forces $b$ to be identically $0$. We will consider further implications of these findings in the next sections. 

\subsection{Ghost and gradient instabilities}
\label{subsec:stability}

The numerical results obtained below show that, for $V=$ const, we have $\rho_{\rm df}+P_{\rm df}=\rho_{\rm df}(1+\omega_{\rm df})>0$ (see Fig.~\ref{fig:ev_p1}). This implies that the dark fluid does not violate the null energy conditions, at least when $V$ is constant. Nevertheless, this does not guarantee that ghost and gradient instabilities are absent. As shown, for instance,  in \cite{Ijjas:2016pad}, a similar mimetic model has unavoidable scalar gradient instabilities while ghost instabilities disappear in certain areas of the parameter space. Our case, however,  is more complicated because of the non-minimal coupling to gravity of the kinetic term $X$, see Eq.~(\ref{lagrangian}). The effects of the non-minimal coupling can be spotted by inspecting Eq.~(\ref{spert}), where the speed of sound $c_{s}^{2}$ is modulated by the factor $(b-c)^{-1}$, but only in the matter sector, i.e. an effective sound speed $c_{s,{\rm eff}}^2=c_s^2/(b-c)$ appears. If $b=0$ and $c>0$ we see that $c_s^2$ and $c_{s,{\rm eff}}^2$ are always opposite in sign.

To shed further light on the behaviour of perturbations, we consider the action to quadratic order in both scalar and tensor perturbations. For the scalar sector a straightforward calculation gives:
\begin{equation}
S^{(2)}_\text{S} =2 (1-a) \int d^4x A^4 H^2 \left[- \frac{1}{c_s^2} \dot{\delta \phi}^2 + \frac{1}{A^2}\left(\partial_k \delta \phi\right) \left(\partial^k \delta \phi\right) + \ldots \right]\,,
\label{quadraticscalar}
\end{equation}
where $\ldots$ stands for terms proportional to $\delta \phi \delta \phi$ and $\delta \phi \dot{\delta \phi}$ which are not important for the stability analysis. For the tensor sector we obtain:
\begin{equation}
S^{(2)}_\text{T} = \frac{1}{4}(1-a) \int d^4x A^3 \left[ \frac{\dot h^2}{c_T^2} - \frac{1}{A^2}\left(\frac{\partial h}{\partial z}\right)^2 \right]\,.
\label{quadratictensor}
\end{equation}

From Eq.~(\ref{quadratictensor}) one sees that in order for tensor perturbations to not suffer from instabilities, we must set $a<1$, so that both terms on the right-hand side of the quadratic action for tensor perturbations appear with the right sign. However, this choice leads, as can be straightforwardly seen from Eq.~(\ref{quadraticscalar}), to gradient instability in the scalar sector and, depending on the sign of $c_{s}^{2}$, also to ghost instability. When $b=a=0$, we recover the same result of~\cite{Ijjas:2016pad}, up to some irrelevant normalisation factors. Thus, as suggested also in this work, the only way to avoid ghost instabilities is to choose $c_{s}^{2}<0$\footnote{In \cite{Ijjas:2016pad} the quantity $(2-3\gamma)/\gamma$ corresponds to our $c_{s}^{-2}$. }. By combining Eqs.~(\ref{friedmann},\ref{sound}) we see that, for $b=0$, $c_{s}^{2}$ and $c$ must have opposite signs. Thus, the conditions $a<1$ and $c>0$ guarantee that the theory is free from ghost instabilities in both the scalar and tensor sectors, although gradient instabilities are still present in the scalar sector.

However, as discussed above, the instability might be tamed by the fact that $c_{s,{\rm eff}}^2=c_s^2/(b-c)>0$ when $c_s^2<0$ and $c>0$ in the limit where $b\rightarrow0$, i.e. the effective sound speed in the presence of matter non-minimally coupled to gravity might actually be positive. A thorough analysis of these perturbed equations goes beyond the scope of this paper but it certainly worth investigating in a follow-up work.

\section{Late-time cosmological evolution}
\label{sec:late_time}

\noindent The main goal of this section is to find solutions mimicking dark matter and/or dark energy in the late Universe, while respecting observational bounds on the speed of scalar and tensor perturbations. To simplify the discussion, we will force the gravitational wave speed $c_T$ to be identically equal to the speed of light, $c_T=1$: as we have seen previously, this implies setting the Lagrangian parameter $b$ to $0$. In other words, a term of the form $\nabla^{\mu}\nabla^{\nu}\phi\nabla_{\mu}\nabla_{\nu}\phi$ is forbidden from appearing in the action, Eq.~(\ref{lagrangian}).

\subsection{Vacuum case}
\label{subsec:vacuum}

Let us begin by considering the idealized case of vacuum, where no cosmological matter is present. Then, the dark fluid density (and pressure) are by definition the effective ones. Then, from the first equation in Eq.~(\ref{friedmann}) combined with Eq.~(\ref{rhoP}) and imposing $b=0$ we get:
\begin{equation}
6H(t)^2 =\frac{4}{4+ 3c - 4a}\left[\frac{C}{A(t)^3}+\frac{3}{A(t)^3}\int^t V(t') A(t')^3 H(t') dt'\right]\,,\label{UNO}
\end{equation}
while the effective Equation of State (EoS) parameter of the Universe, following Eqs.~(\ref{rhoP},\ref{PP}), is given by the following:
\bea
\omega_\text{df}:=\frac{P_\text{df}}{\rho_\text{df}} = \frac{-V}{V+\lambda - 24 a H^2 + (3 c - 12 a)\dot H} \,.\label{eos}
\eea
Now, given a specific form for the scale factor $A(t)$, and therefore a specific form for the Hubble parameter $H(t)$, is possible to reconstruct from Eq.~(\ref{UNO}) the potential $V$ as a function of $t$ and therefore of $\phi$. Moreover, the on-shell form of $\lambda$ can be inferred from Eq.~(\ref{eos}). 

Observations reveal that the Universe today is dominated by dark matter and dark energy. Recent measurements of the CMB temperature and polarization anisotropies and their cross-correlations, in combination with geometrical measurements from Baryon Acoustic Oscillations and Supernovae Type-Ia luminosity distance measurements, suggest that the equation of state of dark energy is extremely close to the the cosmological constant value $\omega=-1$, although small deviations from $-1$ either in the quintessence or the phantom region are still allowed by data~\cite{Planck2015}. In particular, the 68\% confidence level allowed region for the dark energy equation of state is given by~\cite{Planck2015}:
\begin{equation}
-1.0051<\omega_{\rm DE} < -0.961\,,
\label{omega_df_exp_constr}
\end{equation}
Since it is expected that the future evolution of the Universe be dominated by dark energy, let us consider the far-future evolutionary history in our model where we neglect the contribution of the dark matter, setting $C=0$ in Eq.~(\ref{UNO}). In order to describe all the three possible regimes of $\omega_\text{df}$ (cosmological constant-like, quintessence-like, and phantom-like), we will consider the following possibilities:
\begin{align}
A(t)&=\text{e}^{H (t-t_0)} & \omega_\text{df}=-1\,\label{sf1},\\
A(t)&=\left(\frac{t}{t_0}\right)^{\frac{2}{3(1+\omega_{\text{df}})}} & \omega_\text{df}>-1\,\label{sf2},\\
A(t)&=\left(\frac{t^*-t}{t^*-t_0}\right)^{\frac{2}{3(1+\omega_{\text{df}})}} &  \omega_\text{df}<-1\,\label{sf3},
\end{align}
where $t_0$ is the present time for which $A(t_0)= 1$, while $t^*\,, t<t^*$ is the time of the Big Rip~\cite{Caldwell:2003vq} (which emerges within phantom cosmologies, but which can be avoided if a de Sitter Universe is asymptotically reached~\cite{Frampton:2011sp}, or in certain modified gravity theories~\cite{Astashenok:2012tv}). Furthermore, we note that the Hubble parameter $H$ in the case $\omega=-1$ is a constant.

\subsubsection{Potential for $\omega_\text{df} = -1$}

Let us start with the scale factor Eq.~(\ref{sf1}), where $H$ is constant. By choosing the constant potential:
\begin{equation}\label{p1}
V(t) = 2 \Lambda\,,
\end{equation}
where $\Lambda$ is the cosmological constant, from Eq.~(\ref{UNO}) we obtain:
\begin{equation}
6 H^2 = \frac{4}{4+3c-4a}\,(2 \Lambda)\,,
\end{equation}
which is the solution one would expect for a $\Lambda$-dark energy dominated universe (up to the Planck mass rescaling factor). Notice that from Eq.~(\ref{eos}) a constant $\lambda = 24 a H^2$ is needed in order to have $\omega_\text{df} = -1$.

\subsubsection{Potentials for $-1 <\omega_\text{df}$}

Let us now consider an equation of state in the quintessence region. Using the quintessence potential:
\begin{equation}\label{p2}
V(t) = \alpha \left(\frac{t_0}{t}\right)^2 = \alpha \left(\frac{\phi_0}{\phi}\right)^2\,,
\end{equation}
where $\phi_0 = \phi(t_0)$, we recover the scale factor evolution Eq.~(\ref{sf2}). The constant $\alpha$ follows from Eq.~(\ref{UNO}) and reads:
\begin{equation}
\alpha t_0^2 = -\frac{2 \, \omega_\text{df}}{3 (1+\omega_\text{df})^2} (4+3 c - 4 a)\,.
\end{equation}
We see that $\alpha$ is positive when $\omega_\text{df}<0$. More specifically, the constraint  Eq.~(\ref{omega_df_exp_constr}) leads to:
\begin{equation}
420 \, (4 + 3 c - 4 a) \lesssim \alpha t_0^2 < + \infty\,,
\end{equation}
where $\omega_\text{df} = -1$ corresponds to the limit $\alpha \rightarrow + \infty$.

\subsubsection{Potentials for $\omega_\text{df} < -1$}

Finally, let us consider the case where the equation of state is phantom. Similarly to the previous case, we can use a potential of the form:
\begin{equation}\label{p3}
V(t) = \beta \left(\frac{t_*-t_0}{t_*-t}\right)^2 = \beta \left(\frac{\phi_*-\phi_0}{\phi_*-\phi}\right)^2\,,
\end{equation}
where $\phi_* = \phi(t_*)$ and $\phi_0 = \phi(t_0)$, in order to find the scale factor evolution Eq.~(\ref{sf3}). From Eq.~(\ref{UNO}) we find that the constant $\beta$ reads:
\begin{equation}
\beta (t_*-t_0)^2 = -\frac{2 \, \omega_\text{df}}{3 (1+\omega_\text{df})^2} (4+3 c - 4 a)\,,
\end{equation}
and therefore:
\begin{equation}
25800 \, (4 + 3 c - 4 a) \lesssim \beta (t_*-t_0)^2 < + \infty\,,
\end{equation}
where we have used Eq.~(\ref{omega_df_exp_constr}) and $\omega_\text{df} = -1$ corresponds to the limit $\beta \rightarrow + \infty$.

\subsection{Adding radiation, baryons, and dark matter}
\label{subsec:adding}

In this section we numerically solve the system of equations given by the continuity equations for the cosmological matter [Eq.~(\ref{continuity_matter})] and the dark fluid [Eq.~(\ref{continuity_df})], as well as the second Friedmann equation in Eq.~(\ref{friedmann}), using the constant potential $V(t)=2\Lambda$, as in Eq.~(\ref{p1}). 
We consider the action parameter $b=0$ for in order to ensure $c_T=1$ and hence agreement with GW170817/GRB170817A, and we take $a<1$ and $c>0$ (from the requirements on the stability for $b \approx 0$) as free parameters. We further impose that the Planck mass is rescaled to a positive quantity, Eq.~(\ref{condizione}).

We plot the fractional densities defined as:
\begin{equation}\label{frac_density}
\Omega_i(t) = \frac{4}{4+3c-4a}\,\frac{\rho_i(t)}{6 H(t)^2}\,,
\end{equation}
where the index $i$ can correspond to $r$ (radiation), $b$ (baryonic matter) or $df$ (dark fluid), where the dark fluid will include dark energy and dark matter, since in general $C\neq 0$ in Eq.~(\ref{rhodf}). The factor in front of the density is needed if we require the fractional densities of all the components (radiation, baryionic matter and dark fluid) to sum to one at any time $t$. With this definition we expect the $\Omega$'s to not depend on the action parameters $a$ and $c$.

Note that, in order to evolve the system and solve the differential equations, we need to consider some initial conditions. We set $\Omega_r (t_0) = 8 \times 10^{-5}$  and $\Omega_b (t_0) = 0.0486$ (values in agreement with~\cite{Planck2015}), and compute the remaining dark fluid density through $1 - \Omega_r (t_0) -\Omega_b (t_0)$. We evolve the system from a scale factor $A=10^{-5}$ (radiation era) to $A=1$ (present time).

\begin{figure}[htpb]
  \begin{center}
    \includegraphics[width=0.8\textwidth]{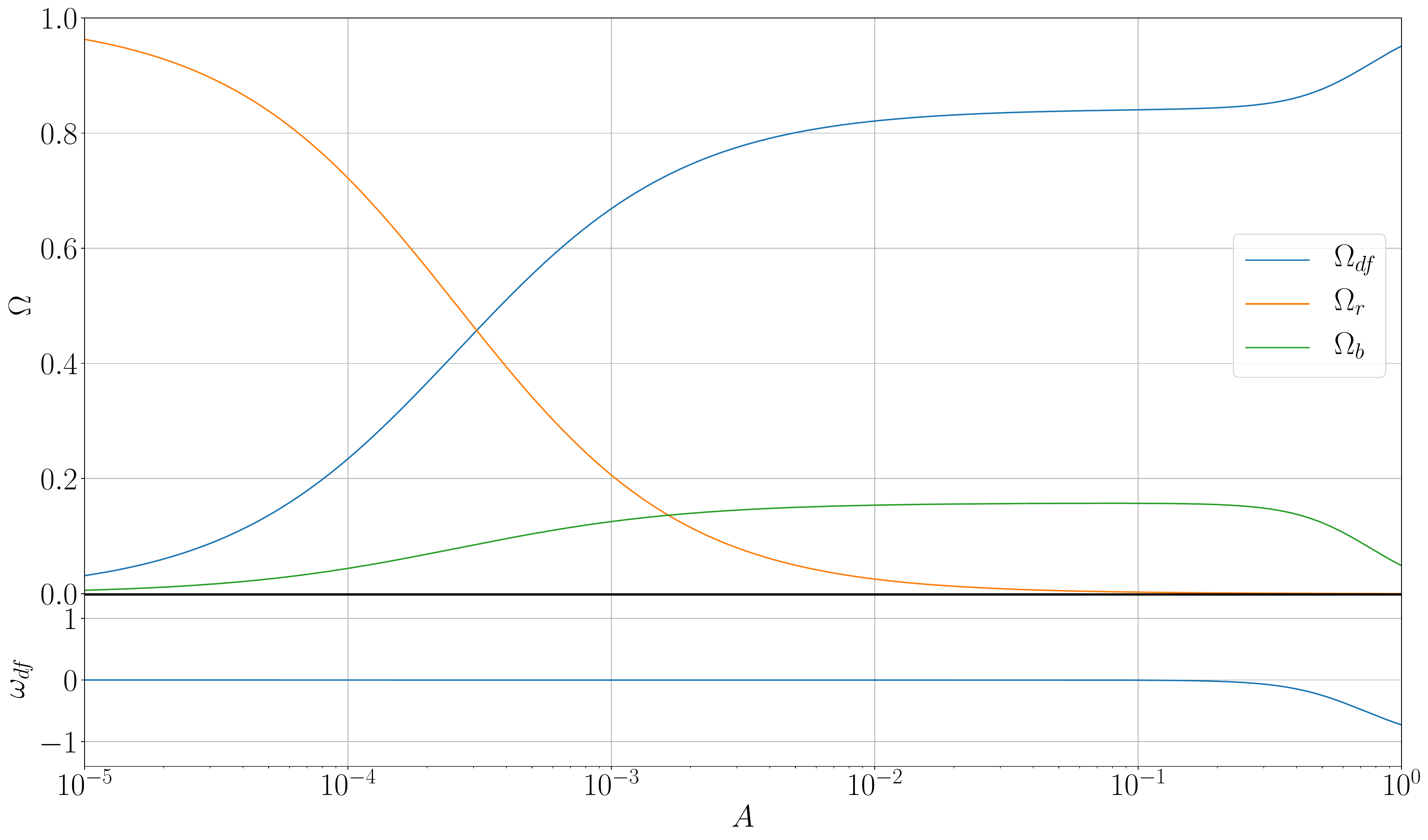}
  \end{center}
  \caption{Plot of the evolution of the fractional densities $\Omega$, defined in equation (\ref{frac_density}), for the cosmological 
  matter components (baryionic matter and radiation) and the dark fluid, and the equation of state parameter $\omega_\text{df}$, as functions of the scale factor $A$. 
  }
  \label{fig:ev_p1}
\end{figure}

The evolution we obtain is depicted in Fig.~\ref{fig:ev_p1}. The fractional densities behave exactly as the ones of $\Lambda$CDM, with the dark fluid corresponding to the cold dark matter and dark energy of $\Lambda$CDM. As already mentioned, we obtain the same fractional densities for every value of $c$ and $a$.

Let us further analyse how varying $a$ and $c$ affects the age of the Universe, which is given by:
\begin{equation}
t_0 \equiv \int_0^1 \, \frac{dA}{A H(A)} = \sqrt{\frac{4+3c-4a}{4}}\int_0^1\, \frac{dA}{A H_0 \sqrt{\frac{1}{6}\sum_i \Omega_i(A)}}\,,
\end{equation}
where $H_0 = H(t_0)$, and we have written the integral as a function of the fractional densities as they do not depend on $a$ and $c$.  Since the factor in front of the integral is $1$ if the action parameters satisfy the condition $4a=3c$ (and therefore also in the case of GR $a=c=0$), we expect the value of the age of the universe to be different from $1$ with a dependence on the action parameters given by
\begin{equation}\label{gamma}
t_0 = \sqrt{\frac{4+3c-4a}{4}} \left.t_0\right|_{4a=3c}\equiv \gamma \left.t_0\right|_{4a=3c}\,.
\end{equation}
Notice that the parameter $\gamma$ is always real, because of the condition imposed in Eq.~(\ref{condizione}). We show some results in Tab.~\ref{tb:age}, where we confirm that Eq.~(\ref{gamma}) is obeyed by our numerical results. The conclusion is that, although the evolution of the fractional densities of the dark fluid corresponds to the one expected in $\Lambda$CDM, the age of the Universe we predict will in general be different unless $4a=3c$. In addition to the constraint on $b$ from the speed of gravitational waves, constraints on the age of the Universe can in principle be used to further constrain the parameters $a$ and $c$.

\begin{figure}[!h]
\begin{center}
\includegraphics[width=1.0\textwidth]{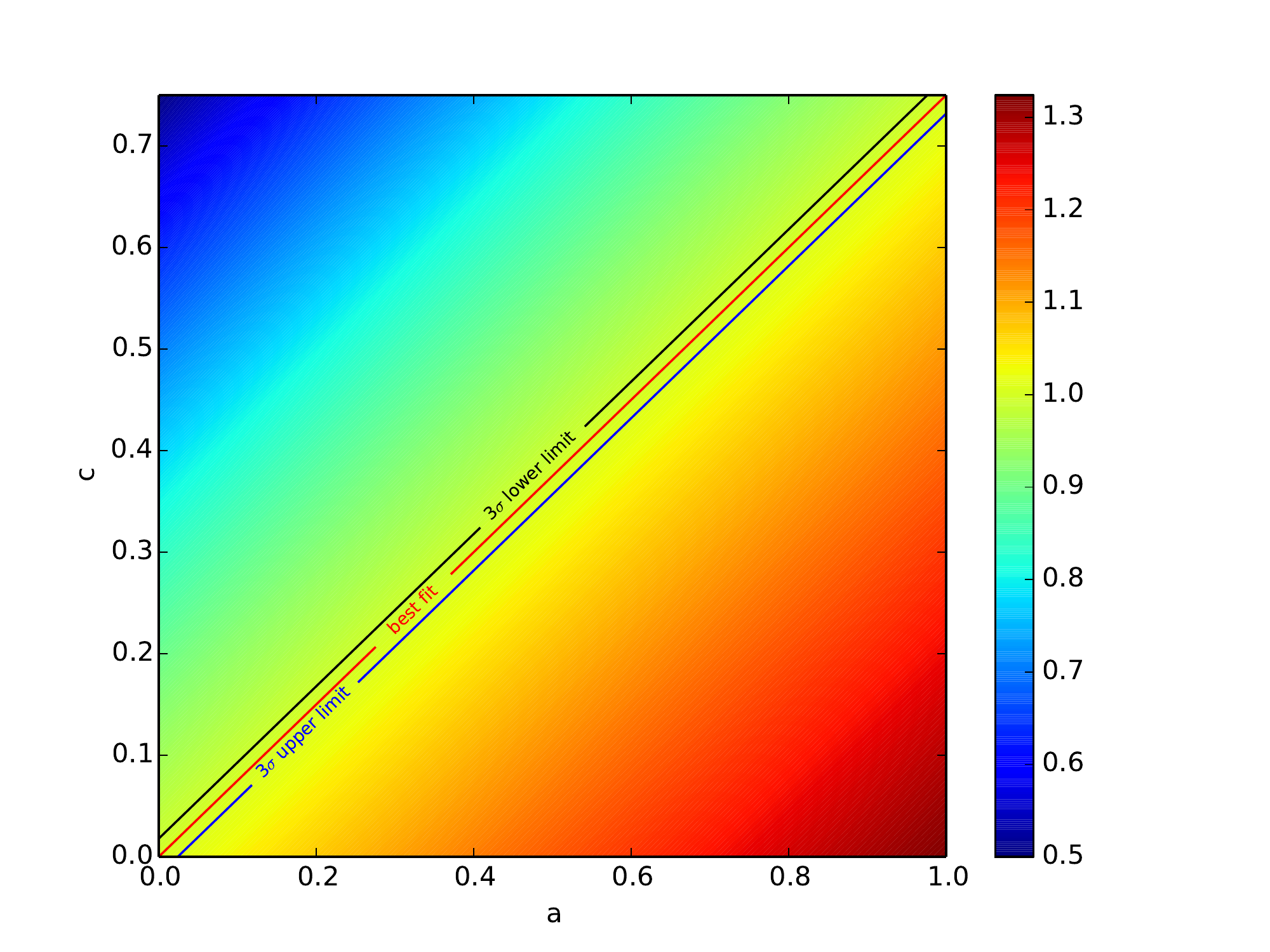}
\end{center}
\caption{Contour-plot of $\gamma=\sqrt{(4+3c-4a)/4}$ in the $a$-$c$ parameter space, focusing for definiteness on the region of parameter space where $a\,,c \lesssim {\cal O}(1)$. The black, red, and blue lines correspond to contours of constant $\gamma=1.009$, $\gamma=1.000$, and $\gamma=0.991$. These values approximately correspond to the $3\sigma$ upper limit, best fit, and $3\sigma$ lower limit on $\gamma$ respectively, given the constraint $\gamma=1.000 \pm 0.003$ which we derive from the $0.3\%$ determination of the age of the Universe from \textit{Planck}~\cite{Planck2015}. Notice that the contours lie along lines of constant $4a-3c$, as expected given the functional form of $\gamma$.}
\label{fig:contourf}
\end{figure}

To get a rough feel for how constraints on the age of the Universe can restrict the viable $a$-$c$ parameter space, we note that the $\approx 0.3\%$ determination of the age of the Universe from the \textit{Planck} satellite~\cite{Planck2015} can in turn be used to constrain $\gamma$, leading to the approximate requirement $\gamma = 1.000 \pm 0.003$. This requirement can be used to set bounds on $a$ and $c$. Focusing for definiteness on the region of parameter space where $a\,,c \lesssim {\cal O}(1)$, in Fig.~\ref{fig:contourf} we show a contour-plot of $\gamma = \sqrt{(4+3c-4a)/4}$ as a function of $a$ and $c$, along with the contours corresponding to $\gamma=1.009$, $\gamma=1.000$, and $\gamma=0.991$. These values approximately correspond to the $3\sigma$ upper limit, best fit, and $3\sigma$ lower limit on $\gamma$ respectively, arising from the constraint $\gamma=1.000 \pm 0.003$. These contours lie along lines of constant values of the linear combination $4a-3c$. Moreover, as expected, we see that limits on the age of the Universe do not constrain the parameters $a$ and $c$ \textit{per se}, but rather the ``orthogonal" linear combination $4a-3c$ (in this sense, the combination $4a-3c$ can be thought of as a principal component of the system), as is clear from the functional form of $\gamma$.

\begin{table}[!h]
\centering
\begin{tabular}{|C{1.5cm}|C{1.5cm}||C{4.3cm}|C{4.3cm}|}
\hline
  $a$&$c$&$t_0$ numerical [Gyr] & $\gamma$\\
  \hline\hline
  $0$&$0$&$13.818$&1\\
  \hline
  $0.75$&$1$&$13.818$&1\\
  \hline
  $0.075$&$0.1$&$13.818$&1\\
  \hline
   $0.1$&$0.1$&$13.644$&$0.99 = 13.644/13.818$\\
   \hline
   $0.5$&$0.1$&$10.478$&$0.76 = 10.478/13.818$\\
  \hline
  $0$&$1$&$18.280$& $1.32 = 18.280/13.818$\\
  \hline
\end{tabular}
\caption{Table of the results for the age of the universe, as function of the action parameters $a$ and $c$. The factor $\gamma$ is defined in equation (\ref{gamma}).}\label{tb:age}
\end{table}

\section{Conclusions}
\label{sec:conclusions}

In this paper we have studied a mimetic model constructed in~\cite{Cognola:2016gjy} by breaking the Horndeski structure of a starting mimetic Horndeski model in order to achieve a non-zero sound speed. We explored the model in light of the recent near-simultaneous detection of GW170817/GRB170817A, which implies that the speed of tensor perturbations $c_T$ is extremely close to the speed of light. In light of this constraint, we then showed how the model can closely mimic the evolution of dark matter and dark energy.

We have found that the stringent constraint on the speed of gravitational waves, equal to the speed of light up to deviations of order 1 part in $10^{15}$~\cite{Monitor:2017mdv}, severely constrains the Lagrangian parameter $b$, which controls the strength of a term of the form $\nabla^{\mu}\nabla^{\nu}\phi\nabla_{\mu}\nabla_{\nu}\phi$ in the action, Eq.~(\ref{lagrangian}). In the limit where we force $c_T$ to be identically equal to the speed of light, $b=0$ is required. We have found that the other two Lagrangian parameters $a$ and $c$ lead to a constant rescaling of the Planck mass, where the unscaled Planck mass of GR is recovered for $3c=4a$. 
The smallness of the parameter $b$, which might lead to a fine-tuning problem, deserves a further comment. When integrated by parts, the relevant term in the action, $\nabla^{\mu}\nabla^{\nu}\phi\nabla_{\mu}\nabla_{\nu}\phi$, leads to a term proportional to $\phi\Box^2\phi$. In~\cite{Saltas:2016awg} (see also~\cite{Brouzakis:2013lla,Saltas:2016nkg}), it was argued that such a term appears when considering 1-loop corrections to the cubic Galileon action. While the analysis of~\cite{Saltas:2016awg} is, not directly applicable to our model, the results tempt us to speculate that the smallness of $b$ might in fact be due to the relevant term being a quantum correction, with the bare parameter being $b=0$. A detailed analysis of the issue, however, is well beyond the scope of this paper, and hence we defer it to future work.

By considering the addition of radiation and baryonic matter, we have numerically solved the modified Friedmann equation and shown that the system can closely mimic an evolutionary history of the Universe consistent with the standard $\Lambda$CDM one at the background level: that is, the mimetic model in question with $b=0$ (in order to comply with constraints from G170817/GRB170817A) can mimic, at the background level, dark matter and dark energy consistently with observations. We have calculated the age of the Universe within the model and find that the $\Lambda$CDM value for this quantity is recovered when $3c=4a$ (which leads to an unscaled Planck mass) as well as for the trivial case where $a=c=0$. Therefore, we expect the approximate relation $3c \simeq 4a$ to hold.\\

A final consideration concerning the stability of the theory is necessary. We have shown that, in order to avoid ghost instabilities, we need  a negative squared sound speed $c_s^2$ together with $c>0$. However, gradient instabilities are still present in the scalar sector but these might be mildened because of the matter field contributions. The stability issue has been studied in many recente papers, see e.g.~\cite{Chaichian:2014qba,Ijjas:2016pad,Firouzjahi:2017txv,Yoshida:2017swb,Hirano:2017zox,Zheng:2017qfs,Cai:2017dyi,
Cai:2017dxl,Takahashi:2017pje,Gorji:2017cai}. While definitive consensus on the matter is yet to be reached, we notice that these issues are likely to affect our model as well, thus mining its theoretical viability. Nonetheless, solutions to these issues have been proposed, involving direct couplings between higher derivatives of the mimetic field and curvature, for instance of the form $f(\Box \phi)$ or $\nabla^{\mu}\nabla^{\nu}R_{\mu \nu}$~\cite{Hirano:2017zox,Zheng:2017qfs,Gorji:2017cai}. Of course, such terms would be expected to modify the prediction for $c_T$ we derived in Eq.~(\ref{tensor}), and could possibly be in conflict with the GW170817/GRB170817A detection. We defer a study of these issues to a separate work.

In conclusion, we have presented a mimetic model which is in perfect agreement with the recent multi-messenger detection of GW170817 and  of GRB170817A, and mimics the evolutionary history of a Universe filled with dark matter and dark energy in agreement with observations at the background level. The model is somewhat appealing in that it appears in the low-energy limit of a well-known candidate theory of quantum gravity, namely Ho\v{r}ava-Lifshitz gravity. A detailed study of cosmological perturbations within the model would be necessary in order to confront it with measurements of the CMB temperature and polarization anisotropies, which constrain many modified gravity models, as well as measurements of the growth of structure, for instance from redshift-space distortions. We defer this issue to future work. In a companion paper~\cite{inprep}, we perform a detailed Bayesian statistical analysis constraining the Lagrangian parameters $a$, $b$, and $c$ in light of the GW170817/GRB170817A detection, therefore allowing for deviations of $c_T$ from the speed of light in agreement with experimental constraints: this represents the \textit{first time} a mimetic model is robustly confronted against observations.

\end{document}